%% file: main.tex
\pdfoutput=1
\documentclass[12pt,a4paper]{article}

\usepackage{ifthen} 
\usepackage{comment}
\usepackage{multirow}
\newboolean{pdflatex}
\setboolean{pdflatex}{true} 

\newboolean{articletitles}
\setboolean{articletitles}{true} 

\newboolean{uprightparticles}
\setboolean{uprightparticles}{false} 

\usepackage{array} 
\usepackage{booktabs} 
\usepackage{multirow} 

\def\paperauthors{LHCb collaboration} 
\def\paperasciititle{Observation of orbitally excited $B_{c}^{+}$ states}
\def\paperkeywords{{High Energy Physics}, {LHCb}} 
\def\papercopyright{\the\year\ CERN for the benefit of the LHCb collaboration} 
\def\paperlicence{CC BY 4.0 licence}
\def\paperlicenceurl{https://creativecommons.org/licenses/by/4.0/}

\newif\ifEnableSectionTOCLinks
\EnableSectionTOCLinksfalse 

\input{preamble}
\usepackage{longtable} 
\input{definitions.tex}

\begin{document}

\renewcommand{\thefootnote}{\fnsymbol{footnote}}
\setcounter{footnote}{1}

\input{title-LHCb-PAPER}

\onecolumn

\renewcommand{\thefootnote}{\arabic{footnote}}
\setcounter{footnote}{0}

\cleardoublepage

\pagestyle{plain} 
\setcounter{page}{1}
\pagenumbering{arabic}



\input{body}

\input{acknowledgements}

\clearpage 
\addcontentsline{toc}{section}{References}
\bibliographystyle{LHCb}
\bibliography{main,standard,LHCb-PAPER,LHCb-CONF,LHCb-DP,LHCb-TDR}

\clearpage 
\input{Authorship_LHCb-PAPER-2025-014}


\end{document}

%% file: preamble.tex
\usepackage[top=1in, bottom=1.25in, left=1in, right=1in]{geometry}

\usepackage{microtype}
\usepackage{lineno}  
\usepackage{xspace} 
\usepackage{caption} 

\usepackage{graphicx}  
\usepackage{color}
\usepackage{colortbl}
\graphicspath{{./figs/}} 

\usepackage{amsmath} 
\usepackage{amssymb}
\usepackage{amsfonts}
\usepackage{upgreek} 

\newcommand*\patchAmsMathEnvironmentForLineno[1]{%
\expandafter\let\csname old#1\expandafter\endcsname\csname #1\endcsname
\expandafter\let\csname oldend#1\expandafter\endcsname\csname
end#1\endcsname
 \renewenvironment{#1}%
   {\linenomath\csname old#1\endcsname}%
   {\csname oldend#1\endcsname\endlinenomath}%
}
\newcommand*\patchBothAmsMathEnvironmentsForLineno[1]{%
  \patchAmsMathEnvironmentForLineno{#1}%
  \patchAmsMathEnvironmentForLineno{#1*}%
}
\AtBeginDocument{%
\patchBothAmsMathEnvironmentsForLineno{equation}%
\patchBothAmsMathEnvironmentsForLineno{align}%
\patchBothAmsMathEnvironmentsForLineno{flalign}%
\patchBothAmsMathEnvironmentsForLineno{alignat}%
\patchBothAmsMathEnvironmentsForLineno{gather}%
\patchBothAmsMathEnvironmentsForLineno{multline}%
\patchBothAmsMathEnvironmentsForLineno{eqnarray}%
}

\usepackage{hyperxmp}

\usepackage[pdftex,
            pdfauthor={\paperauthors},
            pdftitle={\paperasciititle},
            pdfkeywords={\paperkeywords},
            pdfcopyright={Copyright (C) \papercopyright},
            pdflicenseurl={\paperlicenceurl}]{hyperref}

\usepackage[colorinlistoftodos,textsize=scriptsize]{todonotes}

\usepackage[bottom,flushmargin,hang,multiple]{footmisc}

\usepackage[all]{hypcap} 

\input{lhcb-symbols-def} 

\usepackage{cite} 
\usepackage{mciteplus}

%% file: lhcb-symbols-def.tex
\usepackage{xspace} 
\usepackage{upgreek}

\def\lhcb   {\mbox{LHCb}\xspace}

\def\MagUp {\mbox{\em Mag\kern -0.05em Up}\xspace}

\ifthenelse{\boolean{uprightparticles}}%
{

 \def\Ppsi        {\ensuremath{\uppsi}\xspace}

 \def\PDelta      {\ensuremath{\Delta}\xspace}                 
 \def\PXi         {\ensuremath{\Xi}\xspace}                 
 \def\PLambda     {\ensuremath{\Lambda}\xspace}                 
 \def\PSigma      {\ensuremath{\Sigma}\xspace}                 
 \def\POmega      {\ensuremath{\Omega}\xspace}                 
 \def\PUpsilon    {\ensuremath{\Upsilon}\xspace}
 \let\oldPi\Pi
 \def\PPi         {\ensuremath{\oldPi}\xspace}

 \def\PB      {\ensuremath{\mathrm{B}}\xspace}                 
 \def\PD      {\ensuremath{\mathrm{D}}\xspace}                 
 \def\PJ      {\ensuremath{\mathrm{J}}\xspace}                 
 \def\PK      {\ensuremath{\mathrm{K}}\xspace}                 
 \def\Pb      {\ensuremath{\mathrm{b}}\xspace}                 
 \def\Pc      {\ensuremath{\mathrm{c}}\xspace}

 \def\Ps      {\ensuremath{\mathrm{s}}\xspace}

 \def\thebaroffset{0.0em}
}
{

 \def\Ppsi        {\ensuremath{\psi}\xspace}                 
                  
 \mathchardef\PDelta="7101
 \mathchardef\PXi="7104
 \mathchardef\PLambda="7103
 \mathchardef\PSigma="7106
 \mathchardef\POmega="710A
 \mathchardef\PUpsilon="7107
 \mathchardef\PPi="7105
 \def\PB      {\ensuremath{B}\xspace}                 
 \def\PD      {\ensuremath{D}\xspace}                 
 \def\PJ      {\ensuremath{J}\xspace}                 
 \def\PK      {\ensuremath{K}\xspace}                 
 \def\Pb      {\ensuremath{b}\xspace}                 
 \def\Pc      {\ensuremath{c}\xspace}

 \def\Ps      {\ensuremath{s}\xspace}

 \def\thebaroffset{0.18em}
}
\newcommand{\offsetoverline}[2][\thebaroffset]{\kern #1\overline{\kern -#1 #2}}%

\makeatletter
\ifcase \@ptsize \relax
  \newcommand{\miniscule}{\@setfontsize\miniscule{4}{5}}
\or
  \newcommand{\miniscule}{\@setfontsize\miniscule{5}{6}}
\or
  \newcommand{\miniscule}{\@setfontsize\miniscule{5}{6}}
\fi
\makeatother

\DeclareRobustCommand{\optbar}[1]{\shortstack{{\miniscule (\rule[.5ex]{1.25em}{.18mm})}
  \\ [-.7ex] $#1$}}

\def\squark    {{\ensuremath{\Ps}}\xspace}

\def\cquark    {{\ensuremath{\Pc}}\xspace}
\def\cquarkbar {{\ensuremath{\overline \cquark}}\xspace}

\def\bquark    {{\ensuremath{\Pb}}\xspace}
\def\bquarkbar {{\ensuremath{\overline \bquark}}\xspace}

\def\KorKbar {\kern \thebaroffset\optbar{\kern -\thebaroffset \PK}{}\xspace}

\def\D       {{\ensuremath{\PD}}\xspace}

\def\DorDbar {\kern \thebaroffset\optbar{\kern -\thebaroffset \PD}\xspace}

\def\Dp      {{\ensuremath{\D^+}}\xspace}
\def\Dm      {{\ensuremath{\D^-}}\xspace}

\def\DpDm    {\ensuremath{\Dp {\kern -0.16em \Dm}}\xspace}

\def\B       {{\ensuremath{\PB}}\xspace}

\def\BorBbar {\kern \thebaroffset\optbar{\kern -\thebaroffset \PB}\xspace}

\def\Bd      {{\ensuremath{\B^0}}\xspace}

\def\BdorBdbar {\kern \thebaroffset\optbar{\kern -\thebaroffset \Bd}\xspace}

\def\Bs      {{\ensuremath{\B^0_\squark}}\xspace}

\def\BsorBsbar {\kern \thebaroffset\optbar{\kern -\thebaroffset \Bs}\xspace}

\def\Bcp     {{\ensuremath{\B_\cquark^+}}\xspace}

\def\jpsi     {{\ensuremath{{\PJ\mskip -3mu/\mskip -2mu\Ppsi}}}\xspace}

\def\chic     {{\ensuremath{\Pchi_\cquark}}\xspace}

\def\Y#1S{\ensuremath{\PUpsilon{(#1S)}}\xspace}

\def\LorLbar     {\kern \thebaroffset\optbar{\kern -\thebaroffset \PLambda}\xspace}

\def\to                 {\ensuremath{\rightarrow}\xspace}

\def\AT#1     {\ensuremath{A_{\mathrm{T}}^{#1}}\xspace}           

\def\C#1      {\ensuremath{\mathcal{C}_{#1}}\xspace}                       
\def\Cp#1     {\ensuremath{\mathcal{C}_{#1}^{'}}\xspace}                    
\def\Ceff#1   {\ensuremath{\mathcal{C}_{#1}^{\mathrm{(eff)}}}\xspace}        
\def\Cpeff#1  {\ensuremath{\mathcal{C}_{#1}^{'\mathrm{(eff)}}}\xspace}       
\def\Ope#1    {\ensuremath{\mathcal{O}_{#1}}\xspace}                       
\def\Opep#1   {\ensuremath{\mathcal{O}_{#1}^{'}}\xspace}                    

\newcommand{\aunit}[1]{\ensuremath{\text{\,#1}}}       

\newcommand{\tev}{\aunit{Te\kern -0.1em V}\xspace}
\newcommand{\gev}{\aunit{Ge\kern -0.1em V}\xspace}
\newcommand{\mev}{\aunit{Me\kern -0.1em V}\xspace}
\newcommand{\kev}{\aunit{ke\kern -0.1em V}\xspace}
\newcommand{\ev}{\aunit{e\kern -0.1em V}\xspace}
 
\newcommand{\mevc}{\ensuremath{\aunit{Me\kern -0.1em V\!/}c}\xspace}
\newcommand{\gevc}{\ensuremath{\aunit{Ge\kern -0.1em V\!/}c}\xspace}
\newcommand{\mevcc}{\ensuremath{\aunit{Me\kern -0.1em V\!/}c^2}\xspace}
\newcommand{\gevcc}{\ensuremath{\aunit{Ge\kern -0.1em V\!/}c^2}\xspace}

\def\fb   {\ensuremath{\aunit{fb}}\xspace}
\def\invfb   {\ensuremath{\fb^{-1}}\xspace}

\def\gsim{{~\raise.15em\hbox{$>$}\kern-.85em
          \lower.35em\hbox{$\sim$}~}\xspace}
\def\lsim{{~\raise.15em\hbox{$<$}\kern-.85em
          \lower.35em\hbox{$\sim$}~}\xspace}

\def\sqs   {\ensuremath{\protect\sqrt{s}}\xspace}

\def\pt         {\ensuremath{p_{\mathrm{T}}}\xspace}

\def\evtgen     {\mbox{\textsc{EvtGen}}\xspace}

\def\geant      {\mbox{\textsc{Geant4}}\xspace}

\def\photos     {\mbox{\textsc{Photos}}\xspace}

\def\pythia     {\mbox{\textsc{Pythia}}\xspace}

\def\tell1  {TELL1\xspace}
\def\ukl1   {UKL1\xspace}

\newcommand{\lhcborcid}[1]{\href{https://orcid.org/#1}{\hspace*{0.1em}\raisebox{-0.45ex}{\includegraphics[width=1em]{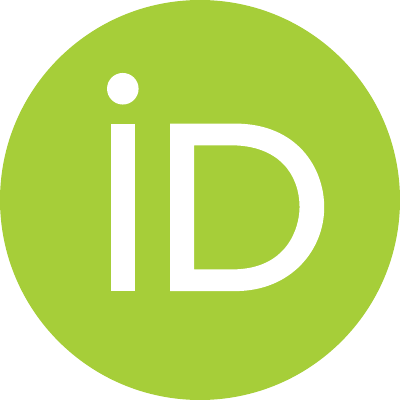}}}}


%% file: definitions.tex
\newcommand{\xx}{\ensuremath{\kern 0.5em }}

\def\Bcstar    {{\ensuremath{\B_\cquark^{*+}}}\xspace}

\usepackage{multirow} 
\usepackage{booktabs} 
\usepackage{rotating}

\def\BcOneP   {B_{c}(1P)^{+}}
\def\BcGamma   {B_{c}^{+}\gamma}

\def\chic{\chi_{c}}


%% file: title-LHCb-PAPER.tex

\begin{titlepage}
\pagenumbering{roman}

\vspace*{-1.5cm}
\centerline{\large EUROPEAN ORGANIZATION FOR NUCLEAR RESEARCH (CERN)}
\vspace*{1.5cm}
\noindent
\begin{tabular*}{\linewidth}{lc@{\extracolsep{\fill}}r@{\extracolsep{0pt}}}
\ifthenelse{\boolean{pdflatex}}
{\vspace*{-1.5cm}\mbox{\!\!\!\includegraphics[width=.14\textwidth]{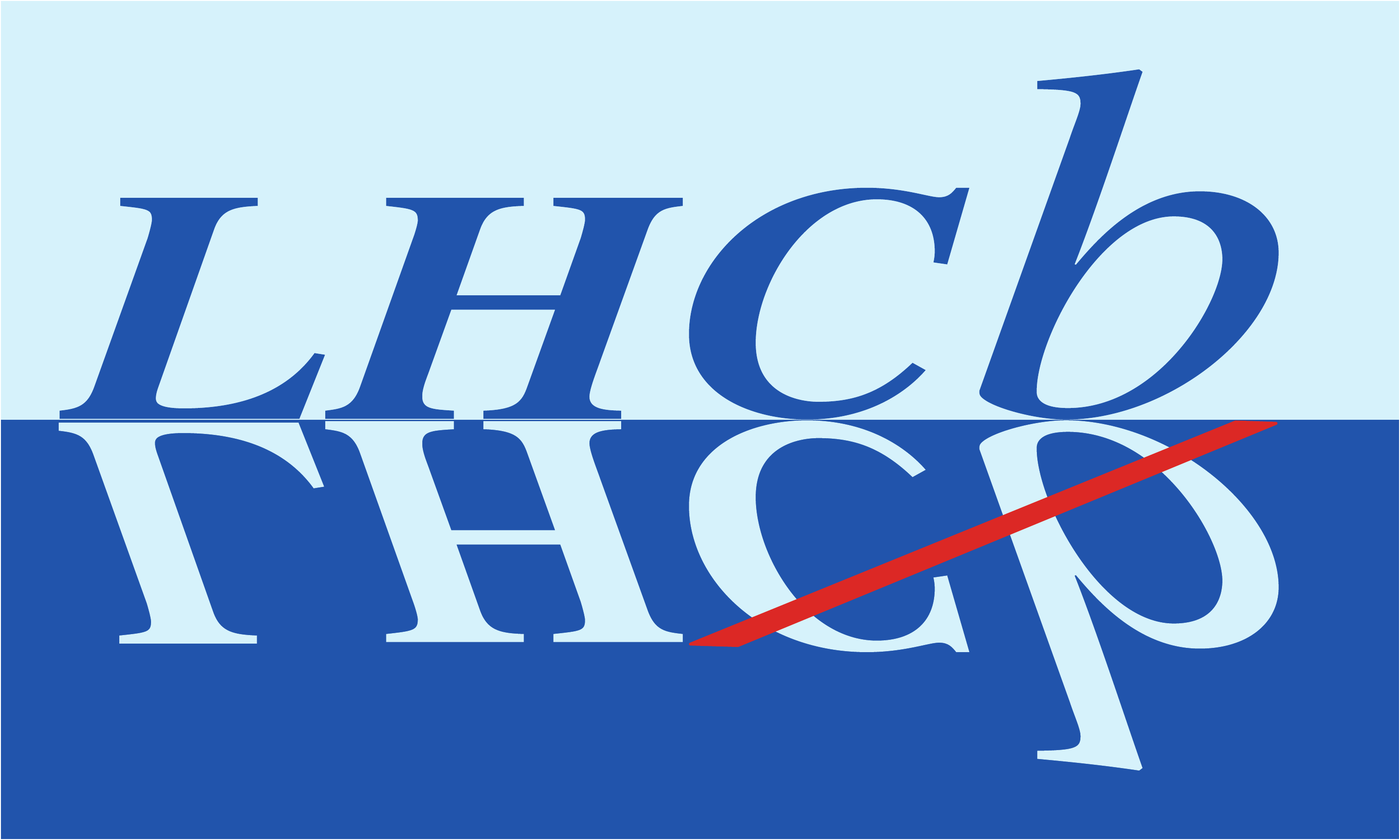}} & &}%
{\vspace*{-1.2cm}\mbox{\!\!\!\includegraphics[width=.12\textwidth]{figs/lhcb-logo.eps}} & &}%
\\
 & & CERN-EP-2025-123 \\  
 & & LHCb-PAPER-2025-014 \\  
 & & 3 December 2025 \\ 
 & & \\
\end{tabular*}

\vspace*{4.0cm}

{\normalfont\bfseries\boldmath\huge
\begin{center}
Observation of orbitally excited \\ $B_c^+$ states
\end{center}
}

\vspace*{2.0cm}

\begin{center}
\paperauthors\footnote{Authors are listed at the end of this Letter.}
\end{center}

\vspace{\fill}

\begin{abstract}
  \noindent
    The observation of a wide peaking structure in the $B_{c}^{+} \gamma$ mass spectrum  
    is reported 
    using proton-proton collision data collected by the LHCb detector at center-of-mass energies of $7$, $8$, and $13~\text{TeV}$, corresponding to a total integrated luminosity of  $9~\text{fb}^{-1}$.
    The statistical significance over the background-only hypothesis exceeds seven standard deviations.
    The width of the observed structure is larger than the expectation from a single-peak hypothesis, and is well described by an effective minimal model consisting of two narrow peaks located at $6704.8 \pm 5.5 \pm 2.8 \pm 0.3~\mathrm{Me\kern -0.1em V\!/}c^2$ and $6752.4 \pm 9.5 \pm 3.1 \pm 0.3~\mathrm{Me\kern -0.1em V\!/}c^2$.
    The uncertainty terms are statistical, systematic, and associated to the knowledge of the $B_{c}^{+}$ mass, respectively.
    The measured peak locations are in line with theoretical predictions for lowest excited $P$-wave $B_{c}^{+}$ states, marking the first observation of orbitally excited beauty-charm mesons and providing important insights into the internal dynamics of hadrons containing two heavy quarks.
\end{abstract}

\vspace*{2.0cm}

\begin{center}
  Published in Phys. Rev. Lett. 135 (2025) 231902.
\end{center}

\vspace{\fill}

{\footnotesize 
\centerline{\copyright~\papercopyright. \href{\paperlicenceurl}{\paperlicence}.}}
\vspace*{2mm}

\end{titlepage}


\newpage
\setcounter{page}{2}
\mbox{~}
%
%
%
%

%% file: body.tex
In the standard model, quantum chromodynamics (QCD) is the theory that governs the strong interaction, which combines quarks and gluons into hadrons. 
In the low-energy regime relevant to hadron formation, the behavior of QCD is inherently nonperturbative, and precise first-principles calculations are notoriously difficult.
Overcoming this challenge remains a key frontier in particle physics research.
The masses of the $\bquark$ and $\cquark$ quarks lie above the nonperturbative QCD scale, allowing for distinctive experimental features and enabling simplified theoretical treatments of heavy hadrons.
As a result, these hadrons play a crucial role for the study of QCD.
Heavy quarkonium systems, including charmonium ($\cquark\cquarkbar$) and bottomonium ($\bquark\bquarkbar$), 
provide a robust testing ground for theoretical models, 
supported by extensive measurements of their mass spectra~\cite{PDG2024,Brambilla:2004wf}.
In contrast, the $\Bcp$ meson family~(charge conjugation is implied throughout the paper), the only hadrons composed of two different heavy quarks ($\bar{b}c$), 
presents both challenges and opportunities.
Its asymmetric quark masses result in a spectrum more intricate than that of  quarkonium systems, 
offering deeper insights into heavy-quark dynamics within hadrons.
Additionally, it serves as a crucial benchmark for validating theoretical predictions, particularly those derived from 
lattice QCD~\cite{Davies:1996gi,Shanahan:1999mv,Allison:2004be,Mathur:2018epb} and various other QCD-based calculations~\cite{
Gershtein:1994dxw,Gupta:1995ps,Fulcher:1998ka,Ebert:2002pp,Godfrey:2004ya,Eichten:2019gig,Li:2019tbn,Wang:2022cxy,Li:2022bre,Li:2023wgq,Hao:2024nqb, 
Chen:1992fq, Zeng:1994vj, El:2005, Wei:2010zza, Dowdall:2012ab, Wang:2012kw,
Karliner:2018bms, 
Chen:2020ecu, Martin-Gonzalez:2022qwd,Nussinov:1983hb,Bertlmann:1979zs,Soni:2017wvy}.
The most precise $\Bcp$ mass predictions within the framework of lattice QCD have been provided by a recent study\mbox{~\cite{Mathur:2018epb}}\hskip0pt, albeit for a subset of the $\BcOneP$ states.
Despite extensive theoretical developments, $\Bcp$ mesons remain experimentally underexplored due to their low production rates compared to charm and beauty hadrons, as their formation requires simultaneous $c\bar{c}$ and $b\bar{b}$ pair production.
Since the discovery of the $\Bcp$ ground state at the Tevatron~\cite{Abe:1998wi,Abe:1998fb}, only the excited $2S$ state, $B_c(2S)^+$, has been observed at the LHC.
It was first reported by the ATLAS experiment as a peaking structure in the $B_c^+\pi^+\pi^-$ mass spectrum~\cite{Aad:2014laa}, 
which was later resolved into two narrower peaks corresponding to the $B_c(2^1S_0)^+$ and $B_c(2^3S_1)^+$ states~\footnote{The spectroscopic notation, $n^{2S+1}L_{J}$, is used, where $n$ gives the radial quantum number, $S$ the sum of the quark spins, $L$ the orbital angular momentum of the quarks, and $J$ the total angular momentum.} by the CMS and LHCb experiments~\cite{CMS:2019uhm,LHCb-PAPER-2019-007}.
With the abundant production of $\bquarkbar\cquark$ states at the LHC, 
the LHCb experiment has significantly advanced the knowledge of their properties, 
providing precise measurements of the $B_c^+$ mass, lifetime, and production rate,
as well as discovering multiple new decay modes,
as summarized in Ref.~\cite{LHCb-PAPER-2023-046}.

Since the total spin of the $\bar{b}$ and $c$ quarks in the $\Bcp$ mesons can be either zero or one, four excited $1P$ states are expected, denoted collectively as $\BcOneP$.
They consist of one axial-vector state with total spin $S=0$ and three states with total spin $S=1$ and total angular momentum $J=0$, $1$, and $2$.
The spin singlet $1^1P_1$ state
is expected to decay into $\Bcp\gamma$, 
while the triplet states $1^3P_0$, $1^3P_1$, and $1^3P_2$ decay into the lowest excited state $1^3S_1$, denoted as $\Bcstar$~\cite{Gouz:2002kk, Godfrey:2004ya,Eichten:2019gig}.
The $\Bcstar$ state subsequently decays into the ground state with the emission of a soft photon that  escapes detection~\cite{Berezhnoy:2013sla,Berezhnoy:2019yei},
hence shifting the peak location of the $\BcOneP$ states by $\delta M$, where $\delta M$ denotes the difference between the $\Bcstar$ and $\Bcp$ masses.
However, due to the mixing of the two axial-vector states, $1^1P_1$ and $1^3P_1$,  into the physical $1P_1$ and $1P_1'$ states, 
they can each decay into both $\Bcp\gamma$ and $\Bcstar\gamma$.
As a result, a total of six mass peaks are expected in the $\Bcp\gamma$ mass spectrum: two peaks at the masses of the $1P_1$ and $1P_1'$ states, and four peaks shifted by $\delta M$ relative to the masses of the $1^3P_0$, $1P_1$, $1P_1'$, and $1^3P_2$ states.
Since the $\Bcstar$ state has not been experimentally observed to date,
the quantity $\delta M$ is unknown and cannot be used to determine the true value of the $\BcOneP$ mass.
The location of the reconstructed peaks in the $M(\Bcp \gamma)-M(\Bcp)$ spectrum is predicted to lie in the range of $[340, 520] \mevcc$~\cite{
Davies:1996gi, 
Gershtein:1994dxw,Gupta:1995ps,Fulcher:1998ka,Ebert:2002pp,Godfrey:2004ya,Eichten:2019gig,Li:2019tbn,Wang:2022cxy,Li:2022bre,Li:2023wgq,Hao:2024nqb}.

This Letter reports the observation of a wide peaking structure in the $\BcGamma$ mass spectrum, 
which can be interpreted as the $\BcOneP$ mesons,
marking the discovery of orbitally excited $\Bcp$ states. 
The study is performed using proton-proton~($pp$) collision data collected with the LHCb detector~\cite{LHCb-DP-2008-001,LHCb-DP-2014-002} from 2011 to 2018 at center-of-mass energies of $\sqs = 7$, $8$, and $13\tev$,
corresponding to integrated luminosities of about $1$, $2$, and $6\invfb$, respectively.
A companion paper Ref.~\cite{LHCb-PAPER-2025-015} is devoted to studying the observed structure from a theory-constrained perspective, where various theoretical models are tested, and the $\BcOneP$ production rate relative to the ground $\Bcp$ state is determined. 

The LHCb detector, a single-arm forward spectrometer covering the pseudorapidity range $2<\eta <5$, is designed for the study of particles containing \bquark or \cquark quarks, with further details provided in Refs.~\cite{LHCb-DP-2008-001,LHCb-DP-2014-002}.
Simulated samples are used to model the signal behavior in $pp$ collisions using \pythia~\cite{Sjostrand:2007gs} with a specific \lhcb configuration~\cite{LHCb-PROC-2010-056}.  
The {\textsc{BcVegPy}} generator~\cite{Chang:2015qea, Chang:2003cq, Chang:2005hq, Wang:2012ah, Wu:2013pya} is employed to produce the $\BcOneP$ and $\Bcp$ states. 
Decays of unstable particles are described by \evtgen~\cite{Lange:2001uf}, in which final-state radiation is generated using \photos~\cite{davidson2015photos}. 
The interaction of the generated particles with the detector, and its response, are implemented using the \geant toolkit~\cite{Allison:2006ve, *Agostinelli:2002hh} as described in Ref.~\cite{LHCb-PROC-2011-006}.

\begin{figure}[t]
 \begin{center}
    \includegraphics[width=0.75\textwidth]{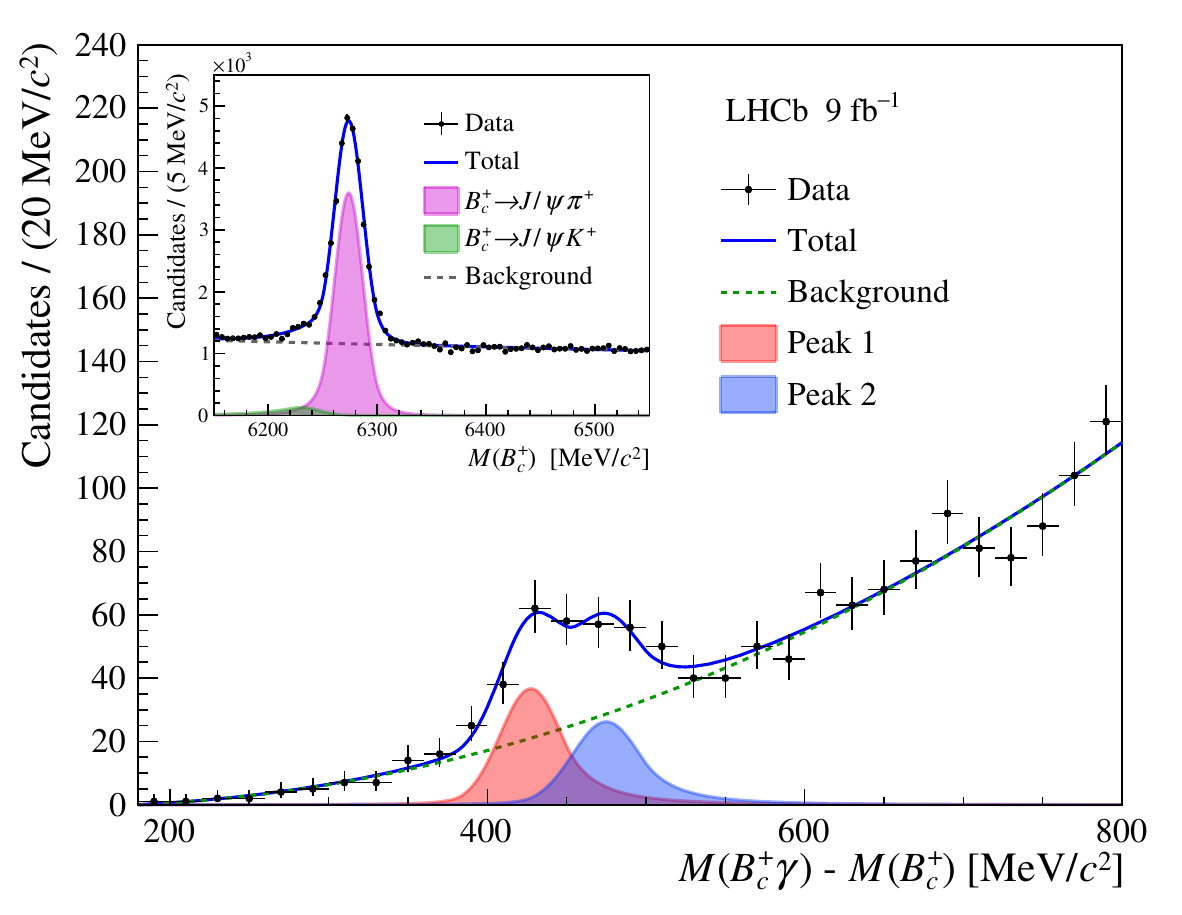}
    \caption{Distribution of the mass difference $M(B_{c}^{+}\gamma)-M(B_{c}^{+})$, where the $B_{c}^{+}$ candidate mass lies within the range $[6240, 6300]~\mathrm{Me\kern -0.1em V\!/}c^2$. The inset displays the mass distribution of the $B_{c}^{+}$ candidates reconstructed from the $B_{c}^{+} \to J/\psi \pi^+$ decay. The fit results are overlaid.}
   \label{fig:massfit}
 \end{center}
\end{figure}

The $\BcOneP$ states are searched for in the $\BcGamma$ mass distribution, with the $\Bcp$ state reconstructed from the $B_c^+ \to J/\psi(\to\mu^+\mu^-)\pi^+$ decay chain.
The mass spectrum of the reconstructed $\Bcp$ candidates is shown in the inset of Fig.~\ref{fig:massfit},
together with the result of an unbinned extended maximum-likelihood fit.
The selection criteria for the photons and the $\Bcp$ candidates, as well as the modeling of each component in the fit, are detailed in Ref.~\cite{LHCb-PAPER-2025-015} and are the same as those used in Refs.~\cite{LHCb-PAPER-2020-003,LHCb-PAPER-2019-007}.
To improve mass resolution, the $\Bcp$ mass  is calculated by constraining the dimuon mass to the known $\jpsi$ mass~\cite{PDG2024} and requiring the $\Bcp$ meson to originate from its associated primary vertex~\cite{Hulsbergen:2005pu}. 
The $\Bcp$ signal yield is found to be $(24.86 \pm 0.24)\times10^3$.
Photons are reconstructed from energy deposits in the electromagnetic calorimeter that are not matched to any track in the event, and combined with $\Bcp$ candidates within the mass range $[6240, 6300] \mevcc$ to form  $\BcOneP$ candidates.
To further improve the signal-to-background ratio, the photon is required to have a large energy component transverse to the beam direction and must satisfy stringent particle-identification requirements based on information from the calorimeter subsystems~\cite{AbellanBeteta:2020amj}.
Background from multiple photon combinations of the same $\Bcp$ meson is suppressed by retaining only the candidate with the photon of highest transverse energy.
The distribution of the mass difference $M(\Bcp \gamma) - M(\Bcp)$ for the selected $\BcOneP$ candidates is shown in Fig.~\ref{fig:massfit}.
A wide peaking structure is present within the predicted mass range for the $\BcOneP$ states.

An unbinned extended maximum-likelihood fit is performed to the distribution of $M(\Bcp \gamma) - M(\Bcp)$.
The signal model is determined from  simulated samples of \mbox{$\BcOneP \to \Bcp \gamma$} decays generated with different $\BcOneP$ masses. The shape is mainly governed by the photon energy scale and resolution.
These are studied using a large sample of $\chic \to \jpsi \gamma$ decays with a photon energy spectrum similar to the signal mode, and the resulting corrections are applied to each simulated sample, as described in Ref.~\cite{LHCb-PAPER-2025-015}.
The correction factors are parametrized as functions of the photon energy and event multiplicity,
and applied to the simulated samples on a per-candidate basis.
The corrected signal shape is modeled by a modified Gaussian function with asymmetric power-law tails, where the Gaussian width is parametrized as a function of the peak location,
and all parameters except the peak location are fixed from simulation.
The natural widths of the $\BcOneP$ states are expected to be only a few hundred $\kev$~\cite{Gershtein:1994dxw,Eichten:1994gt} and are neglected due to their small effect compared to the resolution. 
The combinatorial background is parametrized by a monotonically increasing third-order polynomial function~\cite{karlin1953geometry}.

Using this simple model, the root mean square of the structure is approximately $62\mevcc$, which is significantly larger than the expected value of about $36\mevcc$ under the single-peak hypothesis.
~A better fit quality is obtained using an effective two-peak model, where the locations of both peaks are allowed to vary. The corresponding $p$ value, determined using a Kolmogorov--Smirnov estimation~\cite{smirnov1939estimation}, is found to exceed 90\%. The fit result is shown in Fig.~\ref{fig:massfit}.
The parameters of interest for the minimal two-peak model include the summed signal yields, $N$, and the locations of the two peaks, $M_1$ and $M_2$, which are measured as $N = 182 \pm 25$, $M_1 = 6704.8 \pm 5.5 \, \mevcc$, and $M_2 = 6752.4 \pm 9.5 \, \mevcc$.
The lower-mass peak accounts for $(56 \pm 11)\%$ of the total signal yield.

The signal significance relative to the background-only hypothesis is evaluated using Wilks’ theorem~\cite{Wilks:1938dza} and pseudoexperiments~\cite{Cowan:2010js}. 
Both methods yield consistent results, with the significance exceeding 8 standard deviations,
which remains when alternative signal and background models are considered.
When further accounting for the look-elsewhere effect within the predicted mass region~\cite{Gross:2010qma}, the significance lies above 7 standard deviations.
The $\BcGamma$ mass spectra sourced from different data-taking periods, 
under opposite LHCb dipole magnet polarities, 
and obtained using different selection criteria
are also investigated,
and the presence of a signal structure is confirmed in all of them.
Given the available data and limited mass resolution,
the potential presence of additional peaks, as suggested by theoretical predictions, can neither be confirmed nor excluded. 
In Ref.~\cite{LHCb-PAPER-2025-015}, various theory-constrained six-peak models are used to interpret the observed structure~\mbox{\cite{
Davies:1996gi,
Godfrey:2004ya, 
Ebert:2002pp,
Fulcher:1998ka,
Gershtein:1994dxw,
Gupta:1995ps,
Wang:2022cxy,
Li:2023wgq,
Hao:2024nqb,
Li:2019tbn,
Li:2022bre,
Eichten:2019gig}}.
They are also found to describe the data well, with $p$ values ranging from 15\% to 90\%, which further supports the identification of the peaking structure as $\BcOneP$ states.

Systematic uncertainties in the measurements of the peak locations are detailed below.
The dominant contribution, associated with imperfections in the photon energy correction procedure, is evaluated using the independent decay channel \mbox{$D_{s1}(2460)^{+} \to D_s^+ \gamma$}, which shares similar photon properties with the signal mode. 
Specifically, the correction parameters are applied to the $D_{s1}(2460)^{+}$ simulation, and the largest difference between the $D_{s1}(2460)^{+}$ mass obtained from the corrected simulation and the data is assigned as a systematic uncertainty.
This leads to uncertainties of 2.7 and $3.0\mevcc$ for the lower- and higher-mass structures, respectively.
The uncertainty in the signal shape also arises from unreconstructed photons originating from the $\Bcstar$ decay and imperfect modeling of fully-reconstructed signal decays in the mass fit. 
The former is studied by alternately using the signal shape extracted from $\BcOneP \to \Bcstar(\to \Bcp \gamma) \gamma$ simulation, with the $\Bcstar$ photon unreconstructed.
The latter is assessed by adding another Gaussian function to the original signal model.
The resulting systematic uncertainties from both sources are found to be less than $1 \mevcc$. 
Additionally, the effects of imperfect modeling of the background distributions can slightly alter the results.
The corresponding uncertainty is estimated using an alternative model with a monotonically increasing fourth-order polynomial function, and is found to be negligible.

In summary, using the full $pp$ collision data corresponding to an integrated luminosity of $9\,\mathrm{fb}^{-1}$ collected by the LHCb experiment at center-of-mass energies of 7, 8, and 13\,\tev, 
a wide peaking structure is observed in the $\BcGamma$ mass spectrum with a statistical significance larger than 7 standard deviations.
It is compatible with the lowest excited $P$-wave states of the beauty-charm meson, as verified by fits based on the theoretical predictions from lattice QCD~\cite{Davies:1996gi} and various QCD models~\cite{
Godfrey:2004ya, 
Ebert:2002pp,
Fulcher:1998ka,
Gershtein:1994dxw,
Gupta:1995ps,
Wang:2022cxy,
Li:2023wgq,
Hao:2024nqb,
Li:2019tbn,
Li:2022bre,
Eichten:2019gig} as detailed in Ref.~\cite{LHCb-PAPER-2025-015}.
A minimal model that describes the observed structure well consists of two peaks, with their peak locations, $M_1$ and $M_2$, measured to be
\begin{align*}
    M_1 &= 6704.8 \pm 5.5 \pm 2.8 \pm 0.3 \  \mevcc, \\
    M_2 &= 6752.4 \pm 9.5 \pm 3.1 \pm 0.3 \  \mevcc,
\end{align*}
where the uncertainties are statistical, systematic, and from the limited knowledge of the known $\Bcp$ mass of $6274.5 \pm 0.3 \mevcc$~\cite{PDG2024}.
The two measured peak locations have a statistical correlation of +56\%, while their systematic uncertainties are fully correlated.
This effective model does not account for the expected contributions from the multiple $\BcOneP \to B_c^{(*)+}\gamma$ decays,
the distinction of which requires larger data samples and significant improvements in mass resolution. 
The latter condition can be achieved in the future by using photons converted in the detector material and reconstructed as $e^+e^-$ pairs~\cite{LHCb-PAPER-2013-028,LHCb-PAPER-2014-040},
potentially enabling a measurement of the mass difference between $\Bcstar$ and $\Bcp$ mesons, which is of great theoretical interest and difficult to measure otherwise.
The measurement of the fraction of $\Bcp$ mesons originating from the $\BcOneP$ states, $R$, is detailed in the companion paper Ref.~\cite{LHCb-PAPER-2025-015} using the subset of data collected at $13 \tev$. The fraction $R$ is measured to be $0.20 \pm 0.03 \pm 0.02 \pm 0.03$ within the transverse momentum range $\pt(\Bcp) < 20 \gevc$ and rapidity range $2.0 < y(\Bcp) < 4.5$,
where the uncertainties are statistical, systematic, and related to the choice of theoretical model, respectively.
The obtained value is in good agreement with expectations from nonrelativistic based calculations~\cite{ 
Eichten:2019gig, 
Chang:2005bf,
Chang:2015qea, Chang:2003cq, Chang:2005hq, Wang:2012ah, Wu:2013pya}
, taking different predictions on the feed-down from $B_c(2S)^+$ to $B_c(1P)^+$ states into account~\cite{Godfrey:2004ya,Eichten:2019gig}.

This Letter reports the discovery of orbitally excited beauty-charm states and the measurement of their peak locations using a minimal two-peak model that effectively describes the observed structure.
The results expand known hadron spectroscopy and are crucial to advance our understanding of the internal dynamics of hadrons and the nonperturbative properties of QCD.

\emph{Data availability:} Data associated to the plots in this publication are made available on the CERN Document Server in Ref.~\cite{LHCb-PAPER-2025-014-cds}.

%% file: acknowledgements.tex
\section*{Acknowledgements}
%
%
\noindent
We thank Xing-Gang Wu for valuable discussions on the $\Bcp$ production issues.
We express our gratitude to our colleagues in the CERN
accelerator departments for the excellent performance of the LHC. We
thank the technical and administrative staff at the LHCb
institutes.
We acknowledge support from CERN and from the national agencies:
ARC (Australia);
CAPES, CNPq, FAPERJ and FINEP (Brazil); 
MOST and NSFC (China); 
CNRS/IN2P3 (France); 
BMBF, DFG and MPG (Germany); 
INFN (Italy); 
NWO (Netherlands); 
MNiSW and NCN (Poland); 
MCID/IFA (Romania); 
MICIU and AEI (Spain);
SNSF and SER (Switzerland); 
NASU (Ukraine); 
STFC (United Kingdom); 
DOE NP and NSF (USA).
We acknowledge the computing resources that are provided by ARDC (Australia), 
CBPF (Brazil),
CERN, 
IHEP and LZU (China),
IN2P3 (France), 
KIT and DESY (Germany), 
INFN (Italy), 
SURF (Netherlands),
Polish WLCG (Poland),
IFIN-HH (Romania), 
PIC (Spain), CSCS (Switzerland), 
and GridPP (United Kingdom).
We are indebted to the communities behind the multiple open-source
software packages on which we depend.
Individual groups or members have received support from
Key Research Program of Frontier Sciences of CAS, CAS PIFI, CAS CCEPP, 
Fundamental Research Funds for the Central Universities,  and Sci.\ \& Tech.\ Program of Guangzhou (China);
Minciencias (Colombia);
EPLANET, Marie Sk\l{}odowska-Curie Actions, ERC and NextGenerationEU (European Union);
A*MIDEX, ANR, IPhU and Labex P2IO, and R\'{e}gion Auvergne-Rh\^{o}ne-Alpes (France);
Alexander-von-Humboldt Foundation (Germany);
ICSC (Italy); 
Severo Ochoa and Mar\'ia de Maeztu Units of Excellence, GVA, XuntaGal, GENCAT, InTalent-Inditex and Prog.~Atracci\'on Talento CM (Spain);
SRC (Sweden);
the Leverhulme Trust, the Royal Society and UKRI (United Kingdom).

%% file: Authorship_LHCb-PAPER-2025-014.tex
\centerline
{\large\bf LHCb collaboration}
\begin
{flushleft}
\small
R.~Aaij$^{38}$\lhcborcid{0000-0003-0533-1952},
A.S.W.~Abdelmotteleb$^{57}$\lhcborcid{0000-0001-7905-0542},
C.~Abellan~Beteta$^{51}$\lhcborcid{0009-0009-0869-6798},
F.~Abudin{\'e}n$^{57}$\lhcborcid{0000-0002-6737-3528},
T.~Ackernley$^{61}$\lhcborcid{0000-0002-5951-3498},
A. A. ~Adefisoye$^{69}$\lhcborcid{0000-0003-2448-1550},
B.~Adeva$^{47}$\lhcborcid{0000-0001-9756-3712},
M.~Adinolfi$^{55}$\lhcborcid{0000-0002-1326-1264},
P.~Adlarson$^{84}$\lhcborcid{0000-0001-6280-3851},
C.~Agapopoulou$^{14}$\lhcborcid{0000-0002-2368-0147}
% ...

{\footnotesize \it

% ...
$^{14}$Universit{\'e} Paris-Saclay, CNRS/IN2P3, IJCLab, Orsay, France\\
% ...
$^{38}$Nikhef National Institute for Subatomic Physics, Amsterdam, Netherlands\\
% ...
$^{47}$Instituto Galego de F{\'\i}sica de Altas Enerx{\'\i}as (IGFAE), Universidade de Santiago de Compostela, Santiago de Compostela, Spain\\
% ...
$^{51}$Physik-Institut, Universit{\"a}t Z{\"u}rich, Z{\"u}rich, Switzerland\\
% ...
$^{55}$H.H. Wills Physics Laboratory, University of Bristol, Bristol, United Kingdom\\
% ...
$^{57}$Department of Physics, University of Warwick, Coventry, United Kingdom\\
% ...
$^{60}$School of Physics and Astronomy, University of Glasgow, Glasgow, United Kingdom\\
$^{61}$Oliver Lodge Laboratory, University of Liverpool, Liverpool, United Kingdom\\
% ...
$^{69}$Syracuse University, Syracuse, NY, United States\\
% ...
$^{84}$Department of Physics and Astronomy, Uppsala University, Uppsala, Sweden, associated to $^{60}$\\
% ...
}
\end{flushleft}

%% file: LHCb-DP.bib
@article{LHCb-DP-2014-002,
      author         = "Aaij, R. and others",
      title          = "{LHCb detector performance}",
      collaboration  = "LHCb collaboration",
      journal        = "Int. J. Mod. Phys.",
      volume         = "A30",
      pages          = "1530022",
      doi            = "10.1142/S0217751X15300227",
      year           = "2015",
      eprint         = "1412.6352",
      archivePrefix  = "arXiv",
      primaryClass   = "hep-ex",
      report         = "LHCB-DP-2014-002, CERN-PH-EP-2014-290",
}

@article{LHCb-DP-2008-001,
      author         = "Alves~Jr., A. A. and others",
      title          = "{The \lhcb detector at the LHC}",
      collaboration  = "LHCb collaboration",
      journal        = "JINST",
      volume         = "3",
      pages          = "S08005",
      doi            = "10.1088/1748-0221/3/08/S08005",
      year           = "2008",
      number         = "LHCb-DP-2008-001",
}


%% file: LHCb-PAPER.bib
@article{LHCb-PAPER-2025-015,
      author         = "Aaij, R. and others",
      title          = "{Study of $B_c(1P)^+$ states in the $B_c^+ \gamma$ mass spectrum}",
      collaboration  = "LHCb collaboration",
      report         = "{LHCb-PAPER-2025-015, CERN-EP-2025-122}",
      eprint         = "2507.02142",
      archivePrefix  = "arXiv",
      primaryClass   = "hep-ex",
      journal        = "{Phys. Rev.}",
      volume         = "D112",
      year           = "2025",
      pages          = "112003",
      doi            = "10.1103/1d49-q8h4"
}

@article{LHCb-PAPER-2023-046,
      author         = "Aaij, R. and others",
      title          = "{Observation of the $\Bcp \to \jpsi \pip \piz$ decay}",
      collaboration  = "LHCb collaboration",
      report         = "{LHCb-PAPER-2023-046, CERN-EP-2024-019}",
      eprint         = "2402.05523",
      archivePrefix  = "arXiv",
      primaryClass   = "hep-ex",
      journal           = "{JHEP}",
      volume = "04",
      pages  = "151",
      year     = "2024",
      doi = "10.1007/JHEP04(2024)151",
}

@article{LHCb-PAPER-2020-003,
      author         = "Aaij, R. and others",
      title          = "{Precision measurement of the \Bc meson mass}",
      collaboration  = "LHCb collaboration",
      report         = "{LHCb-PAPER-2020-003, CERN-EP-2020-048}",
      eprint         = "2004.08163",
      archivePrefix  = "arXiv",
      primaryClass   = "hep-ex",
      year           = "2020",
      journal        = "JHEP",
      volume         = "07",
      pages          = "123",
      doi            = "10.1007/JHEP07(2020)123",
}

@article{LHCb-PAPER-2019-007,
      author         = "Aaij, R. and others",
      title          = "{Observation of an excited \Bc state}",
      collaboration  = "LHCb collaboration",
      report         = "{LHCb-PAPER-2019-007 CERN-EP-2019-050}",
      eprint         = "1904.00081",
      archivePrefix  = "arXiv",
      primaryClass   = "hep-ex",
      year           = "2019",
      journal        = "Phys. Rev. Lett.",
      volume         = "122",
      pages          = "232001",
      doi            = "10.1103/PhysRevLett.122.232001",
}

@article{LHCb-PAPER-2014-040,
      author         = "Aaij, R. and others",
      title          = "{Measurement of the $\chi_b(3P)$ mass and of the relative 
                         rate of $\chi_{b1}(1P)$ and $\chi_{b2}(1P)$ production}",
      collaboration  = "LHCb collaboration",
      journal        = "JHEP",
      volume         = "10",
      pages          = "088",
      doi            = "10.1007/JHEP10(2014)088",
      year           = "2014",
      eprint         = "1409.1408",
      archivePrefix  = "arXiv",
      primaryClass   = "hep-ex",
      report         = "LHCb-PAPER-2014-040-CERN_PH_EP_2014_206",
}

@article{LHCb-PAPER-2013-028,
      author         = "Aaij, R. and others",
      title          = "{Measurement of the relative rate of prompt $\chiczero$, 
                         $\chicone$ and $\chictwo$ production at \mbox{$\sqs=$7~\tev}}",
      collaboration  = "LHCb collaboration",
      journal        = "JHEP",
      volume         = "10",
      pages          = "115",
      doi            = "10.1007/JHEP10(2013)115",
      year           = "2013",
      eprint         = "1307.4285",
      archivePrefix  = "arXiv",
      primaryClass   = "hep-ex",
      report         = "CERN-PH-EP-2013-114 LHCb-PAPER-2013-028",
}


%% file: main.bib
@article{Dowdall:2012ab,
      author         = "Dowdall, R. J. and Davies, C. T. H. and Hammant, T. C.
                        and Horgan, R. R.",
      title          = "{Precise heavy-light meson masses and hyperfine
                        splittings from lattice QCD including charm quarks in the
                        sea}",
      journal        = "Phys. Rev.",
      volume         = "D86",
      year           = "2012",
      pages          = "094510",
      doi            = "10.1103/PhysRevD.86.094510",
      eprint         = "1207.5149",
      archivePrefix  = "arXiv",
      primaryClass   = "hep-lat",
      SLACcitation   = "%%CITATION = ARXIV:1207.5149;%%"
}

@article{Aad:2014laa,
      author         = "Aad, Georges and others",
      title          = "{Observation of an excited $B_c^\pm$ meson state with the
                        ATLAS detector}",
      collaboration  = "ATLAS collaboration",
      journal        = "Phys. Rev. Lett.",
      volume         = "113",
      year           = "2014",
      pages          = "212004",
      doi            = "10.1103/PhysRevLett.113.212004",
      eprint         = "1407.1032",
      archivePrefix  = "arXiv",
      primaryClass   = "hep-ex",
      reportNumber   = "CERN-PH-EP-2014-137",
      SLACcitation   = "%%CITATION = ARXIV:1407.1032;%%"
}

@article{Shanahan:1999mv,
    author = "Shanahan, H. P. and Boyle, P. and Davies, C. T. H. and Newton, H.",
    collaboration = "UKQCD",
    title = "{A nonperturbative calculation of the mass of the $B_c$}",
    eprint = "hep-lat/9902025",
    archivePrefix = "arXiv",
    reportNumber = "UTCCP-P-53, GUTPA-98-12-02, EDINBURGH-98-23",
    doi = "10.1016/S0370-2693(99)00325-1",
    journal = "Phys. Lett.",
    volume = "B453",
    pages = "289--294",
    year = "1999"
}

@Article{Gouz:2002kk,
author = "Gouz, I. P. and Kiselev, V. V. and Likhoded, A. K. and
Romanovsky, V. I. and Yushchenko, O. P.",
title = "{Prospects for the $B_c$ studies at LHCb}",
journal = "Phys. Atom. Nucl.",
volume = "67",
year = "2004",
pages = "1559-1570",
eprint = "hep-ph/0211432",
archivePrefix = "arXiv",
doi = "10.1134/1.1788046",
SLACcitation = "%%CITATION = HEP-PH/0211432;%%"
}

@Article{Chang:2003cq,
author = "Chang, Chao-Hsi and Driouichi, Chafik and Eerola, Paula and
Wu, Xing Gang",
title = "{{\textsc{BcVegPy}}: An event generator for hadronic production of the
$B_c$ meson}",
journal = "Comput. Phys. Commun.",
volume = "159",
year = "2004",
pages = "192-224",
eprint = "hep-ph/0309120",
archivePrefix = "arXiv",
doi = "10.1016/j.cpc.2004.02.005",
SLACcitation = "%%CITATION = HEP-PH/0309120;%%"
}

@Article{Chang:2005hq,
author = "Chang, Chao-Hsi and Wang, Jian-Xiong and Wu, Xing-Gang",
title = "{{\textsc{BcVegPy}}2.0: An upgrade version of the generator {\textsc{BcVegPy}}
with an addendum about hadroproduction of the $P$-wave $B_c$
states}",
journal = "Comput. Phys. Commun.",
volume = "174",
year = "2006",
pages = "241-251",
eprint = "hep-ph/0504017",
archivePrefix = "arXiv",
doi = "10.1016/j.cpc.2005.09.008",
SLACcitation = "%%CITATION = HEP-PH/0504017;%%"
}

@article{Wang:2012ah,
    author = "Wang, Xian-You and Wu, Xing-Gang",
    title = "{A~trick to improve the~efficiency of generating 
    unweighted $\PB_\cquark$~events from {\textsc{BcVegPy}}}",
    eprint = "1108.2442",
    archivePrefix = "arXiv",
    primaryClass = "hep-ph",
    reportNumber = "SLAC-PUB-15007",
    doi = "10.1016/j.cpc.2011.08.024",
    journal = "Comput. Phys. Commun.",
    volume = "183",
    pages = "442--445",
    year = "2012"
}

@article{Wu:2013pya,
    author = "Wu, Xing-Gang",
    editor = "Wang, Jianxiong",
    title = "{{\textsc{BcVegPy}} and {\sc{GenXicc}} 
   for the~hadronic production of the~doubly heavy mesons and baryons}",
    eprint = "1307.3344",
    archivePrefix = "arXiv",
    primaryClass = "hep-ph",
    doi = "10.1088/1742-6596/523/1/012042",
    journal = "J. Phys. Conf. Ser.",
    volume = "523",
    pages = "012042",
    year = "2014"
}

@Article{Chen:1992fq,
author = "Chen, Yu-Qi and Kuang, Yu-Ping",
title = "{Improved QCD-motivated heavy quark potentials with
explicit $\Lambda_{\overline{\rm MS}}$ dependence}",
journal = "Phys. Rev.",
volume = "D46",
year = "1992",
pages = "1165-1171",
doi = "10.1103/PhysRevD.46.1165",
SLACcitation = "%%CITATION = PHRVA,D46,1165;%%",
extraPrefix    = "Erratum",
extraVolume    = "D47",
extraPages     = "350",
extraYear      = "1993",
extraDoi       = "10.1103/PhysRevD.47.350",
}

@Article{Eichten:1994gt,
author = "Eichten, Estia J. and Quigg, Chris",
title = "{Mesons with beauty and charm: spectroscopy}",
journal = "Phys. Rev.",
volume = "D49",
year = "1994",
pages = "5845-5856",
eprint = "hep-ph/9402210",
archivePrefix = "arXiv",
doi = "10.1103/PhysRevD.49.5845",
SLACcitation = "%%CITATION = HEP-PH/9402210;%%"
}

@Article{Gupta:1995ps,
author = "Gupta, Suraj N. and Johnson, James M.",
title = "{$B_c$ spectroscopy in a quantum-chromodynamic potential
model}",
journal = "Phys. Rev.",
volume = "D53",
year = "1996",
pages = "312-314",
eprint = "hep-ph/9511267",
archivePrefix = "arXiv",
doi = "10.1103/PhysRevD.53.312",
SLACcitation = "%%CITATION = HEP-PH/9511267;%%"
}

@Article{Fulcher:1998ka,
author = "Fulcher, Lewis P.",
title = "{Phenomenological predictions of the properties of the $B_c$
system}",
journal = "Phys. Rev.",
volume = "D60",
year = "1999",
pages = "074006",
eprint = "hep-ph/9806444",
archivePrefix = "arXiv",
doi = "10.1103/PhysRevD.60.074006",
SLACcitation = "%%CITATION = HEP-PH/9806444;%%"
}

@Article{Ebert:2002pp,
author = "Ebert, D. and Faustov, R. N. and Galkin, V. O.",
title = "{Properties of heavy quarkonia and $B_c$ mesons in the
relativistic quark model}",
journal = "Phys. Rev.",
volume = "D67",
year = "2003",
pages = "014027",
eprint = "hep-ph/0210381",
archivePrefix = "arXiv",
doi = "10.1103/PhysRevD.67.014027",
SLACcitation = "%%CITATION = HEP-PH/0210381;%%"
}

@Article{Godfrey:2004ya,
author = "Godfrey, Stephen",
title = "{Spectroscopy of $B_c$ mesons in the relativized quark
model}",
journal = "Phys. Rev.",
volume = "D70",
year = "2004",
pages = "054017",
eprint = "hep-ph/0406228",
archivePrefix = "arXiv",
doi = "10.1103/PhysRevD.70.054017",
SLACcitation = "%%CITATION = HEP-PH/0406228;%%"
}

@Article{Davies:1996gi,
author = "Davies, C. T. H. and others",
title = "{$B_c$ spectroscopy from lattice QCD}",
journal = "Phys. Lett.",
volume = "B382",
year = "1996",
pages = "131-137",
eprint = "hep-lat/9602020",
archivePrefix = "arXiv",
doi = "10.1016/0370-2693(96)00650-8",
SLACcitation = "%%CITATION = HEP-LAT/9602020;%%"
}

@Article{Brambilla:2004wf,
author = "Brambilla, N. and others",
collaboration = "Quarkonium Working Group",
title = "{Heavy quarkonium physics}",
year = "2004",
eprint = "hep-ph/0412158",
archivePrefix = "arXiv",
SLACcitation = "%%CITATION = HEP-PH/0412158;%%"
}

@article{Allison:2004be,
    author = "Allison, Ian F. and Davies, Christine T. H. and Gray, Alan and Kronfeld, Andreas S. and Mackenzie, Paul B. and Simone, James N.",
    collaboration = "HPQCD, Fermilab Lattice and UKQCD",
    title = "{Mass of the $B_c$ meson in three-flavor lattice QCD}",
    eprint = "hep-lat/0411027",
    archivePrefix = "arXiv",
    reportNumber = "FERMILAB-PUB-04-349-T",
    doi = "10.1103/PhysRevLett.94.172001",
    journal = "Phys. Rev. Lett.",
    volume = "94",
    pages = "172001",
    year = "2005"
}

@Article{Zeng:1994vj,
author = "Zeng, J. and Van Orden, J. W. and Roberts, W.",
title = "{Heavy mesons in a relativistic model}",
journal = "Phys. Rev.",
volume = "D52",
year = "1995",
pages = "5229-5241",
eprint = "hep-ph/9412269",
archivePrefix = "arXiv",
doi = "10.1103/PhysRevD.52.5229",
SLACcitation = "%%CITATION = HEP-PH/9412269;%%"
}

@Article{Abe:1998wi,
author = "Abe, F. and others",
collaboration = "CDF collaboration",
title = "{Observation of the $B_c$ meson in $p\bar{p}$ collisions at
$\sqrt{s} = 1.8$ TeV}",
journal = "Phys. Rev. Lett.",
volume = "81",
year = "1998",
pages = "2432-2437",
eprint = "hep-ex/9805034",
archivePrefix = "arXiv",
doi = "10.1103/PhysRevLett.81.2432",
SLACcitation = "%%CITATION = HEP-EX/9805034;%%"
}

@Article{Abe:1998fb,
author = "Abe, F. and others",
collaboration = "CDF collaboration",
title = "{Observation of $B_c$ mesons in $p\bar{p}$ collisions at
$\sqrt{s} = 1.8$ TeV}",
journal = "Phys. Rev.",
volume = "D58",
year = "1998",
pages = "112004",
eprint = "hep-ex/9804014",
archivePrefix = "arXiv",
doi = "10.1103/PhysRevD.58.112004",
SLACcitation = "%%CITATION = HEP-EX/9804014;%%"
}

@article{Wei:2010zza,
author         = "Wei, Ke-Wei and Guo, Xin-Heng",
title          = "{Mass spectra of doubly heavy mesons in Regge
	phenomenology}",
journal        = "Phys. Rev.",
volume         = "D81",
year           = "2010",
pages          = "076005",
doi            = "10.1103/PhysRevD.81.076005",
SLACcitation   = "%%CITATION = PHRVA,D81,076005;%%"
}

@article{El:2005,
title = "{Radiative decays of ${B}_{c}$ mesons in a Bethe--Salpeter model}",
author = {Abd El-Hady, A. and Spence, J. R. and Vary, J. P.},
journal = {Phys. Rev.},
volume = {D71},
issue = {3},
pages = {034006},
numpages = {5},
year = {2005},
month = {Feb},
publisher = {American Physical Society},
doi = {10.1103/PhysRevD.71.034006},
url = {http://link.aps.org/doi/10.1103/PhysRevD.71.034006},
eprint         = "hep-ph/0603139",
archivePrefix  = "arXiv",
primaryClass   = "hep-ph",
SLACcitation   = "%%CITATION = HEP-PH/0603139;%%"
}

@article{Chang:2005bf,
   author         = "Chang, Chao-Hsi and Qiao, Cong-Feng and Wang, Jian-Xiong
	and Wu, Xing-Gang",
   title          = "{Color-octet contributions to P-wave $B_c$ meson
	hadroproduction}",
journal        = "Phys. Rev.",
volume         = "D71",
year           = "2005",
pages          = "074012",
doi            = "10.1103/PhysRevD.71.074012",
eprint         = "hep-ph/0502155",
archivePrefix  = "arXiv",
primaryClass   = "hep-ph",
SLACcitation   = "%%CITATION = HEP-PH/0502155;%%"
}

@article{Berezhnoy:2013sla,
   author         = "Berezhnoy, Aleksander and Likhoded, Anatolii",
   title          = "{The observation possibility of $B_c$ excitations at
	LHC}",
   booktitle      = "{Proceedings, 21st International Workshop on High Energy
   Physics and Quantum Field Theory (QFTHEP 2013): St.
								   Petersburg, Russia, June 23-30, 2013}",
   journal        = "PoS QFTHEP2013",
   year           = "2013",
   pages          = "051",
   eprint         = "1307.5993",
   archivePrefix  = "arXiv",
   primaryClass   = "hep-ph",
   SLACcitation   = "%%CITATION = ARXIV:1307.5993;%%"
}

@article{Cowan:2010js,
      author         = "Cowan, Glen and Cranmer, Kyle and Gross, Eilam and
                        Vitells, Ofer",
      title          = "{Asymptotic formulae for likelihood-based tests of new
                        physics}",
      journal        = "Eur. Phys. J.",
      volume         = "C71",
      year           = "2011",
      pages          = "1554",
      doi            = "10.1140/epjc/s10052-011-1554-0",
      eprint         = "1007.1727",
      archivePrefix  = "arXiv",
      primaryClass   = "physics.data-an",
      SLACcitation   = "%%CITATION = ARXIV:1007.1727;%%",
      extraPrefix    = "Erratum",
      extraVolume    = "C73",
      extraPages     = "2501",
      extraYear      = "2013",
      extraDoi       = "10.1140/epjc/s10052-013-2501-z",
}

@article{Gross:2010qma,
      author         = "Gross, Eilam and Vitells, Ofer",
      title          = "{Trial factors for the look elsewhere effect in high
                        energy physics}",
      journal        = "Eur. Phys. J.",
      volume         = "C70",
      year           = "2010",
      pages          = "525-530",
      doi            = "10.1140/epjc/s10052-010-1470-8",
      eprint         = "1005.1891",
      archivePrefix  = "arXiv",
      primaryClass   = "physics.data-an",
      SLACcitation   = "%%CITATION = ARXIV:1005.1891;%%"
}

@article{Chang:2015qea,
      author         = "Chang, Chao-Hsi and Wang, Xian-You and Wu, Xing-Gang",
      title          = "{{\textsc{BcVegPy}}2.2: a newly upgraded version for hadronic
                        production of the meson $\Bc$ and its excited states}",
      journal        = "Comput. Phys. Commun.",
      volume         = "197",
      year           = "2015",
      pages          = "335-338",
      doi            = "10.1016/j.cpc.2015.07.015",
      eprint         = "1507.05176",
      archivePrefix  = "arXiv",
      primaryClass   = "hep-ph",
      SLACcitation   = "%%CITATION = ARXIV:1507.05176;%%"
}

@article{CMS:2019uhm,
    author = "Sirunyan, Albert M and others",
    collaboration = "CMS",
    title = "{Observation of two excited $B_{c}^{+}$ states and measurement of the $B_{c}^{+}(2S)$ mass in pp collisions at $\sqrt{s} =$ 13 TeV}",
    eprint = "1902.00571",
    archivePrefix = "arXiv",
    primaryClass = "hep-ex",
    reportNumber = "CMS-BPH-18-007, CERN-EP-2019-014",
    doi = "10.1103/PhysRevLett.122.132001",
    journal = "Phys. Rev. Lett.",
    volume = "122",
    number = "13",
    pages = "132001",
    year = "2019"
}

@article{Eichten:2019gig,
    author = "Eichten, Estia J. and Quigg, Chris",
    title = "{Mesons with beauty and charm: new horizons in spectroscopy}",
    eprint = "1902.09735",
    archivePrefix = "arXiv",
    primaryClass = "hep-ph",
    reportNumber = "FERMILAB-PUB-19/075-T, FERMILAB--PUB--19/075--T, FERMILAB-PUB-19-270-T",
    doi = "10.1103/PhysRevD.99.054025",
    journal = "Phys. Rev.",
    volume = "D99",
    number = "5",
    pages = "054025",
    year = "2019"
}

@book{karlin1953geometry,
  title={Geometry of moment spaces},
  author={Karlin, Samuel and Shapley, Lloyd S},
  volume={12},
  year={1953},
  publisher={American Mathematical Soc.}
}

@article{AbellanBeteta:2020amj,
    author = "Abell\'an Beteta, Carlos and others",
    title = "{Calibration and performance of the LHCb calorimeters in Run 1 and 2 at the LHC}",
    eprint = "2008.11556",
    archivePrefix = "arXiv",
    primaryClass = "physics.ins-det",
    reportNumber = "CERN-LHCb-DP-2020-001",
    month = "8",
    year = "2020"
}

@article{Li:2023wgq,
    author = "Li, Xue-Jian and Li, Yu-Shuai and Wang, Fu-Lai and Liu, Xiang",
    title = "{Spectroscopic survey of higher-lying states of $B_c$ meson family}",
    eprint = "2308.07206",
    archivePrefix = "arXiv",
    primaryClass = "hep-ph",
    doi = "10.1140/epjc/s10052-023-12237-9",
    journal = "Eur. Phys. J.",
    volume = "C83",
    number = "11",
    pages = "1080",
    year = "2023"
}

@article{Hao:2024nqb,
    author = "Hao, Wei and Zhu, Ruilin",
    title = "{Beauty-charm meson family with coupled channel effects and their strong decays}",
    eprint = "2402.18898",
    archivePrefix = "arXiv",
    primaryClass = "hep-ph",
    doi = "10.1088/1674-1137/ad75f5",
    journal = "Chin. Phys.",
    volume = "C48",
    number = "12",
    pages = "123101",
    year = "2024"
}

@article{Wang:2022cxy,
    author = "Wang, Guo-Li and Wang, Tianghong and Li, Qiang and Chang, Chao-Hsi",
    title = "{The mass spectrum and wave functions of the B$_{c}$ system}",
    eprint = "2201.02318",
    archivePrefix = "arXiv",
    primaryClass = "hep-ph",
    doi = "10.1007/JHEP05(2022)006",
    journal = "JHEP",
    volume = "05",
    pages = "006",
    year = "2022"
}

@article{smirnov1939estimation,
  title={On the estimation of the discrepancy between empirical curves of distribution for two independent samples},
  author={Smirnov, Nikolai V},
  journal={Bull. Math. Univ. Moscou},
  volume={2},
  number={2},
  pages={3--14},
  year={1939}
}

@article{Gershtein:1994dxw,
    author = "Gershtein, S. S. and Kiselev, V. V. and Likhoded, A. K. and Tkabladze, A. V.",
    title = "{ $B_c$ spectroscopy}",
    eprint = "hep-ph/9406339",
    archivePrefix = "arXiv",
    reportNumber = "IFVE-94-51",
    doi = "10.1103/PhysRevD.51.3613",
    journal = "Phys. Rev.",
    volume = "D51",
    pages = "3613--3627",
    year = "1995"
}

@article{Chen:2020ecu,
    author = "Chen, Muyang and Chang, Lei and Liu, Yu-Xin",
    title = "{$B_c$ meson spectrum via Dyson--Schwinger equation and Bethe--Salpeter equation approach}",
    eprint = "2001.00161",
    archivePrefix = "arXiv",
    primaryClass = "hep-ph",
    doi = "10.1103/PhysRevD.101.056002",
    journal = "Phys. Rev.",
    volume = "D101",
    number = "5",
    pages = "056002",
    year = "2020"
}

@article{Karliner:2018bms,
    author = "Karliner, Marek and Rosner, Jonathan L.",
    title = "{Scaling of P-wave excitation energies in heavy-quark systems}",
    eprint = "1808.07869",
    archivePrefix = "arXiv",
    primaryClass = "hep-ph",
    reportNumber = "EFI 18-11, TAUP 3033/18",
    doi = "10.1103/PhysRevD.98.074026",
    journal = "Phys. Rev.",
    volume = "D98",
    number = "7",
    pages = "074026",
    year = "2018"
}

@article{Li:2019tbn,
    author = {Li, Qi and Liu, Ming-Sheng and Lu, Long-Sheng and L\"u, Qi-Fang and Gui, Long-Cheng and Zhong, Xian-Hui},
    title = "{Excited bottom-charmed mesons in a nonrelativistic quark model}",
    eprint = "1903.11927",
    archivePrefix = "arXiv",
    primaryClass = "hep-ph",
    doi = "10.1103/PhysRevD.99.096020",
    journal = "Phys. Rev.",
    volume = "D99",
    number = "9",
    pages = "096020",
    year = "2019"
}

@article{Martin-Gonzalez:2022qwd,
    author = "Mart\'\i{}n-Gonz\'alez, B. and Ortega, P. G. and Entem, D. R. and Fern\'andez, F. and Segovia, J.",
    title = "{Toward the discovery of novel $B_c$ states: radiative and hadronic transitions}",
    eprint = "2205.05950",
    archivePrefix = "arXiv",
    primaryClass = "hep-ph",
    doi = "10.1103/PhysRevD.106.054009",
    journal = "Phys. Rev.",
    volume = "D106",
    number = "5",
    pages = "054009",
    year = "2022"
}

@article{Li:2022bre,
    author = "Li, Ting-Yan and Tang, Long and Fang, Zheng-yuan and Wang, Chao-hui and Pang, Cheng-qun and Liu, Xiang",
    title = "{Higher states of the $B_c$ meson family}",
    eprint = "2204.14258",
    archivePrefix = "arXiv",
    primaryClass = "hep-ph",
    doi = "10.1103/PhysRevD.108.034019",
    journal = "Phys. Rev.",
    volume = "D108",
    number = "3",
    pages = "034019",
    year = "2023"
}

@article{Wang:2012kw,
    author = "Wang, Zhi-Gang",
    title = "{Analysis of the vector and axialvector $B_c$ mesons with QCD sum rules}",
    eprint = "1203.6252",
    archivePrefix = "arXiv",
    primaryClass = "hep-ph",
    doi = "10.1140/epja/i2013-13131-7",
    journal = "Eur. Phys. J.",
    volume = "A49",
    pages = "131",
    year = "2013"
}

@article{Mathur:2018epb,
    author = "Mathur, Nilmani and Padmanath, M. and Mondal, Sourav",
    title = "{Precise predictions of charmed-bottom hadrons from lattice QCD}",
    eprint = "1806.04151",
    archivePrefix = "arXiv",
    primaryClass = "hep-lat",
    reportNumber = "TIFR/TH/18-09, TIFR-TH-18-09",
    doi = "10.1103/PhysRevLett.121.202002",
    journal = "Phys. Rev. Lett.",
    volume = "121",
    number = "20",
    pages = "202002",
    year = "2018"
}

@article{Berezhnoy:2019yei,
    author = "Berezhnoy, A. V. and Belov, I. N. and Likhoded, A. K. and Luhinsky, A. V.",
    title = "{$B_c$ excitations at LHC experiments}",
    eprint = "1904.06732",
    archivePrefix = "arXiv",
    primaryClass = "hep-ph",
    doi = "10.1142/S0217732319503310",
    journal = "Mod. Phys. Lett.",
    volume = "A34",
    number = "40",
    pages = "1950331",
    year = "2019"
}

@article{Bertlmann:1979zs,
    author = "Bertlmann, R. A. and Martin, Andre",
    title = "{Inequalities on heavy quark-antiquark systems}",
    reportNumber = "CERN-TH-2772",
    doi = "10.1016/0550-3213(80)90279-5",
    journal = "Nucl. Phys. ",
    volume = "B168",
    pages = "111--136",
    year = "1980"
}

@article{Nussinov:1983hb,
    author = "Nussinov, Shmuel",
    title = "{Mass inequalities in {QCD}}",
    reportNumber = "Print-83-0958 (MARYLAND), MdDP-TR-84-64",
    doi = "10.1103/PhysRevLett.52.966",
    journal = "Phys. Rev. Lett.",
    volume = "52",
    pages = "966",
    year = "1984"
}

@article{Soni:2017wvy,
    author = "Soni, N. R. and Joshi, B. R. and Shah, R. P. and Chauhan, H. R. and Pandya, J. N.",
    title = "{$Q\bar{Q}$ ( $Q\in \{b, c\}$ ) spectroscopy using the Cornell potential}",
    eprint = "1707.07144",
    archivePrefix = "arXiv",
    primaryClass = "hep-ph",
    doi = "10.1140/epjc/s10052-018-6068-6",
    journal = "Eur. Phys. J. ",
    volume = "C78",
    number = "7",
    pages = "592",
    year = "2018"
}

@misc{LHCb-PAPER-2025-014-cds,
    author = "",
    howpublished = {\url{https://cds.cern.ch/record/2937508}}
}


%% file: standard.bib
@article{PDG2024,
     author    = "Navas, S. and others",
    collaboration = "Particle Data Group",
     title     = "{\href{http://pdg.lbl.gov/}{Review of particle physics}}",
     journal   = "Phys. Rev.",
     year = {2024},
     number = {8},
     volume      = "D110",
     pages     = "030001",
     doi = "10.1103/PhysRevD.110.030001"
}

@article{LHCb-PROC-2011-006,
  author="Clemencic, M and others",
  title="{The \lhcb simulation application, {\textsc{Gauss}}: Design, evolution and experience}",
  journal="J. Phys. Conf. Ser.",
  volume={331},
  pages={032023},
  doi={10.1088/1742-6596/331/3/032023},
  year={2011},
}

@article{LHCb-PROC-2010-056,
      author         = "Belyaev, I. and others",
      title          = "{Handling of the generation of primary events
                         in {\textsc{Gauss}}, the LHCb simulation framework}",
      journal="J. Phys. Conf. Ser.",
      volume={331},
      pages={032047},
      doi={10.1088/1742-6596/331/3/032047},
      year={2011},
}

@article{Sjostrand:2007gs,
      author         = {Sj\"{o}strand, Torbj\"{o}rn and Mrenna, Stephen and
                        Skands, Peter"},
      title          = "{A brief introduction to {\textsc{Pythia}} 8.1}",
      journal        = "Comput. Phys. Commun.",
      volume         = "178",
      pages          = "852-867",
      doi            = "10.1016/j.cpc.2008.01.036",
      year           = "2008",
      eprint         = "0710.3820",
      archivePrefix  = "arXiv",
      primaryClass   = "hep-ph",
      reportNumber   = "CERN-LCGAPP-2007-04, LU-TP-07-28,
                        FERMILAB-PUB-07-512-CD-T",
}

@Article{Agostinelli:2002hh,
     author    = "Agostinelli, S. and others",
 collaboration = "Geant4 collaboration",
     title     = "{{\textsc{Geant4}}: A simulation toolkit}",
     journal   = "Nucl. Instrum. Meth.",
     volume    = "A506",
     year      = "2003",
     pages     = "250",
     doi       = "10.1016/S0168-9002(03)01368-8",
}

@article{Allison:2006ve,
      author         = "Allison, John and Amako, K. and Apostolakis, J. and
                        Araujo, H. and Dubois, P.A. and others",
 collaboration = "Geant4 collaboration",
      title          = "{{\textsc{Geant4}} developments and applications}",
      journal        = "IEEE Trans.Nucl.Sci.",
      volume         = "53",
      pages          = "270",
      doi            = "10.1109/TNS.2006.869826",
      year           = "2006",
      reportNumber   = "SLAC-PUB-11870",
}

@Article{Lange:2001uf,
     author    = "Lange, D. J.",
     title     = "{The {\textsc{EvtGen}} particle decay simulation package}",
     journal   = "Nucl. Instrum. Meth.",
     volume    = "A462",
     year      = "2001",
     pages     = "152-155",
     doi       = "10.1016/S0168-9002(01)00089-4",
}

@article{davidson2015photos,
	author="Davidson, N. and Przedzinski, T. and Was, Z.",
	title = "{{\textsc{Photos}} interface in C++: Technical and physics documentation}",
      	eprint={1011.0937},
      	archivePrefix={arXiv},
      	primaryClass={hep-ph},
	journal = {Comp. Phys. Comm.},
	volume = {199},
	pages = {86},
	year = {2016},
	doi = {https://doi.org/10.1016/j.cpc.2015.09.013},
}

@article{Hulsbergen:2005pu,
      author         = "Hulsbergen, Wouter D.",
      title          = "{Decay chain fitting with a Kalman filter}",
      journal        = "Nucl. Instrum. Meth.",
      volume         = "A552",
      pages          = "566-575",
      doi            = "10.1016/j.nima.2005.06.078",
      year           = "2005",
      eprint         = "physics/0503191",
      archivePrefix  = "arXiv",
      primaryClass   = "physics",
}

@article{Wilks:1938dza,
      author         = "Wilks, S. S.",
      title          = "{The large-sample distribution of the likelihood ratio
                        for testing composite hypotheses}",
      journal        = "Ann. Math.Stat.",
      volume         = "9",
      pages          = "60-62",
      doi            = "10.1214/aoms/1177732360",
      year           = "1938",
}

@Misc{mciteplus,
  author = 	 {Shell, Michael},
  title = 	 {Mciteplus: Enhanced multicitations},
  howpublished = {\href{http://www.michaelshell.org/tex/mciteplus/} {http://www.michaelshell.org/tex/mciteplus/}},
}
